\documentclass[prc,floatfix,groupedaddress,nofootinbib,showpacs,showkeys,preprintnumbers,
amsmath,amssymb,amsfonts,superscriptaddress,widetable]{revtex4-1}
\usepackage{bm}
\usepackage{mathrsfs}
\usepackage{amssymb}
\usepackage{amsmath}
\usepackage{amsfonts}
\usepackage{graphicx}
\usepackage{array}
\usepackage[usenames,x11names]{xcolor}

\newcommand{\psub}[1]{{\bf p}_{\raisebox{-2pt}{$\scriptstyle#1$}}}
\newcommand{\sigmasub}[1]{{\sigma}_{\raisebox{-2pt}{$\scriptscriptstyle\hspace{-1pt}\!\!#1$}}}
\newcommand{\rhosub}[1]{\rho_{\raisebox{-0.75pt}{\tiny{#1}}}}
\newcommand{\mc}[1]{\mathcal{#1}}

\begin{document}
\title{Information and statistics: a new paradigm in theoretical nuclear physics\hfill}
\author{J. Piekarewicz}\email{jpiekarewicz@fsu.edu}
\author{Wei-Chia Chen}\email{wc09c@my.fsu.edu}
\affiliation{Department of Physics, Florida State University,
               Tallahassee, FL 32306, USA}
\author{F.J. Fattoyev}\email{farrooh.fattoyev@tamuc.edu}
\affiliation{Department of Physics and Astronomy, Texas A\&M
               University-Commerce, Commerce, TX 75429, USA}
\date{\today}
\begin{abstract}
 Theoretical predictions of physical observables often involve extrapolations
 to regions that are poorly constrained by laboratory experiments and
 astrophysical observations. Without properly quantified theoretical errors,
 such model predictions are of very limited utility. In this contribution we
 use \emph{maximum-likelihood} estimation to compute theoretical errors
 and assess correlations between physical observables. We illustrate the
 power and elegance of these methods using examples of both pedagogical
 and realistic interest. In particular, we implement a gaussian approximation
 to the likelihood function to develop a new relativistic effective interaction
 constrained by ground-state properties of finite nuclei, their monopole
 response, and masses of neutron stars.
\end{abstract}

\smallskip

\maketitle

\section{Introduction}
\label{Introduction}

Quantum Chromodynamics (QCD) is the fundamental theory of the strong
interactions. As such, all strong-interaction phenomena should in
principle be described in terms of QCD. In practice, however, such a
task remains a daunting one even after more than 40 years since its
discovery. However, the properties and underlying symmetries of QCD
(e.g., chiral symmetry) may serve to dictate the form of an
\emph{effective theory} to describe nuclear phenomena. To date, the
best---and perhaps only---alternative to account for the richness of
nuclear phenomena---ranging from the properties of finite nuclei to
the structure of neutron stars---is the density functional theory
(DFT). The enormous advantage of DFT is that it shifts the focus
from the complicated many-body wave-function to the much simpler and
physically intuitive one-body density\,\cite{Kohn:1999}. Moreover,
Kohn and Sham have shown how the one-body density may be obtained
from a variational problem that reduces to the solution of a set of
mean-field-like equations\,\cite{Kohn:1965}. Although the form of
the Kohn-Sham potential is highly reminiscent of the Hartree (or
Hartree-Fock) potential used as a starting point in many-body
calculations, the constants that parametrize the potential are
directly fitted to \emph{many-body observables} rather than to
two-body data. In this manner some of the complicated many-body
dynamics originating from exchange and correlation effects gets
implicitly encoded in the empirical constants of the model. Given
that the \emph{nuclear energy density functional} (EDF) must be
constrained from experimental data, the determination of the model
parameters must rely on the optimization of a suitably defined
\emph{objective function} (e.g., $\chi^{2}$).

The conventional approach to the calibration of the EDF consists of
first minimizing the objective function and then validating the
model against observables not included in the fit. Often the
predictions involve observables that remain within the region of
validity of the model. However, occasionally these predictions
involve extrapolations well outside such region---as in the case of
neutron stars. How should one assess those predictions? Frankly,
without reliable theoretical uncertainties, such predictions become
of limited value. Quoting the quintessential American author Mark
Twain: \emph{``I've come loaded with statistics, for I've noticed
that a man can't prove anything without statistics".} Unfortunately,
such wit and wisdom has yet to be fully appreciated by the
theoretical community. Indeed---and as so aptly put by the Editors
of the Physical Review A---\emph{``it is all too often the case that
the numerical results are presented without uncertainty
estimates"}\,\cite{PhysRevA.83.040001}.

Fortunately, the need to apply the same high standards to
theoretical predictions that one has applied for many years to
laboratory measurements and astrophysical observations is gaining
significant momentum. For example, the successful UNEDF
collaboration advocates the \emph{`search for a microscopic theory
that both predicts and provides well-quantified theoretical
uncertainties'}\,\cite{UNEDF,Kortelainen:2010hv}. Further, within
the last few years a significant number of publication have
addressed---not only the critical role of theoretical errors---but
also the wealth of information that is contained in the study of
statistical correlations among observables\,\cite{Reinhard:2010wz,
Fattoyev:2011ns,Fattoyev:2012rm,Piekarewicz:2012pp,Reinhard:2012vw,
Erler:2012qd,Reinhard:2013fpa,Fattoyev:2014pja}. The present focus
issue is a true testament to our commitment to continue the growth
in this area.

Among the critical issues that \emph{information and statistics} can
address is how fast does the ``optimal model"---namely, the one that
minimizes the objective function---deteriorate as one moves away
from the minimum. That is, are there certain directions in parameter
space that are too ``soft" such that moving along such directions
does not result in a significant deterioration of the fit? If so, is
this evidence that some sectors of the EDF are poorly constrained by
the choice of physical observables? And if this is the case, which
additional observables should be added to better constrain the
model? Finally, if such observables are hard to determine, are there
other ones that are both strongly correlated to and more readily
accessible than the original ones? Although we will provide answers
to all these questions in what follows, a well-known topical example
may serve as an illustration. The example involves a quantity of
critical importance in constraining the equation of state of
neutron-rich matter, namely, the slope of the symmetry energy $L$.
At present however, well-measured nuclear observables---such as
binding energies and charge radii---are insufficient to constrain
the \emph{isovector} sector of the EDF. This implies that
predictions of $L$ using accurately-calibrated EDFs are accompanied
by large theoretical errors. That is, $L$ is sensitive to a linear
combination of parameters that defines a very soft direction in
parameter space\,\cite{Fattoyev:2011ns,Fattoyev:2012rm}. Moreover,
the task is further complicated by the fact that $L$, being a bulk
property of infinite nuclear matter, can not be measured in the
laboratory. One must then find an alternative to $L$ that is both
experimentally accessible and strongly correlated to it. An
observable that has been shown to satisfy both of these requirements
is the neutron-skin thickness of
${}^{208}$Pb\,\cite{Brown:2000,Furnstahl:2001un,RocaMaza:2011pm}. In
essence, we want to infer by how much our knowledge of $L$ will
improve from an accurate measurement of the neutron-skin thickness
of ${}^{208}$Pb. The power of inference may be demonstrated with an
example that dates all the way back to the 18th century. It appears
that Laplace used Bayes' theorem to decide which astronomical
problems to work on. In particular, from the data available at the
end of the 18th century, Laplace estimated the mass of Saturn to
be 3,512 times smaller than the mass of the Sun---and gave a
better than 99\% probability that the actual mass lies within 1\%
of that value. As of the latest compilation, the mass of Saturn
[(1/3,499)$M_{\odot}$] deviates from Laplace's prediction by a
mere 0.3\%.

The manuscript has been organized as follows. In
Sec.\,\ref{Formalism} we develop the formalism required to address
the role of information and statistics in theoretical nuclear
physics. In particular, we frame the discussion in terms of the
\emph{likelihood function} and obtain analytic results in the
gaussian limit, that corresponds to quadratic oscillations around
the $\chi^{2}$ minimum. In Sec.\,\ref{Results} we illustrate the
main concepts developed in the formalism by using both pedagogical
and realistic examples. In particular, we discuss the calibration of
a new relativistic EDF for the description of finite nuclei and
neutron stars. Finally, our conclusions and outlook are presented in
Sec.\,\ref{Conclusions}.

\section{Formalism}
\label{Formalism}

Models of nuclear structure are characterized by a number of
parameters. The aim of any calibration procedure is to determine the
set of model parameters that best describes a given set of
experimental data. Such an optimal parameter set is obtained by
demanding that a suitably defined {\sl objective function} be
minimized. In the traditional least-squares-fit
method\,\cite{Brandt:1999,Bevington2003}, the objective function to
be minimized is the sum of the squares of the deviations between the
experimental data and the predictions of the model. Assuming that
the model contains a total of $F$ free parameters
${\bf p}\!=\!(p_{1},\ldots,p_{F})$ and that a number $N\!\gg\!F$ of
experimental observables $\mc{O}_{n}^{\rm (exp)}$
($n\!=\!1,\ldots,N$) is used for the calibration, the objective function
is defined as\,\cite{Reinhard:2010wz,
Fattoyev:2011ns,Fattoyev:2012rm,Dobaczewski:2014jga}:
\begin{equation}
 \chi^{2}({\bf p}) \equiv \sum_{n=1}^{N}
 \frac{\Big(\mc{O}_{n}^{\rm (th)}({\bf p})-
               \mc{O}_{n}^{\rm (exp)}\Big)^{2}}
 {\Delta\mc{O}_{n}^{2}} \,,
 \label{ChiSquare}
\end{equation}
where $\mc{O}_{n}^{\rm (th)}({\bf p})$ is the theoretical prediction
and ${\Delta\mc{O}_{n}}$ represents the adopted errors, which in
general contain both experimental and theoretical contributions.
Whereas the definition of $\chi^{2}$ is standard and several powerful
techniques for finding its minimum value already exist, our aim is to
highlight the critical role of theoretical uncertainties.
Indeed, as articulated in an editorial published in the Physical Review
A\,\cite{PhysRevA.83.040001}, theoretical predictions are now
expected to be accompanied by meaningful uncertainty estimates.
In particular, the need for ``theoretical error bars'' becomes critical
whenever models calibrated in certain domain are used to extrapolate
into uncharted regions.

There are two sources of errors that will be addressed here:
statistical and systematic. The estimation of statistical errors
associated with a given model, although fairly novel in nuclear
theory, is relatively straightforward to implement. In particular,
the statistical errors and the underlying correlations between
observables are revealed by the structure of the parameter space
around the region of the optimal model\,\cite{Fattoyev:2011ns}. In
contrast, and as already addressed by Dobaczewski, Nazarewicz, and
Reinhard in this focus issue\,\cite{Dobaczewski:2014jga}, there are
various contributions to the systematic error that are difficult to
quantify. For example, how does one quantify whether the underlying
model is robust? That is, whether the model is capable of
reproducing the rich dynamics displayed by nuclear systems so that
the correlations suggested by the model be trustworthy.  Further, is
the experimental data used to define the objective function
adequate, or are there large sections of parameter space that remain
unconstrained? Finally, and perhaps hardest, what should one assume
for the adopted errors ${\Delta\mc{O}_{n}}$.  The experimental
contribution to the error is relatively simple to assess as it
consists of the quoted error in the measurement of ${O}_{n}^{\rm
(exp)}$.  However, one must also include a theoretical component to
the error that is highly uncertain. That a theoretical component to
the error must exist is clear, otherwise the objective function
becomes dominated by the observable with the smallest experimental
error. For instance, in the particular case of ${}^{208}$Pb, its
binding energy has been determined to an astonishing precision of
about 1 part per million, its charge radius to about 2 parts in
10,000, and the centroid energy of its monopole resonance to about
2\%.  However, what is not clear is the precise value that one must
adopt for the theoretical contribution to the error. Unfortunately,
there is no foolproof answer to this question---although
Ref.\,\cite{Dobaczewski:2014jga} provides a useful guiding
principle. Ultimately, the optimum choice of theoretical errors
necessary involves some ``trial and error".

\subsection{The Likelihood Function}
\label{Likelihood}

A concept that will be used extensively throughout this manuscript and
one that plays a critical role in statistics is the \emph{likelihood
function}. The likelihood function is closely related to the objective
function defined in Eq.\,(\ref{ChiSquare}) and represents, as its name
indicates, the likelihood that a set of model parameters ${\bf p}$
reproduces the given experimental data ${O}_{n}^{\rm (exp)}$. The
likelihood function is defined as follows:
\begin{equation}
 {\mc L}\Big({\bf p}|\mc{O}_{n}^{\rm (exp)}\Big)
 \equiv {\mc L}({\bf p}) =
 \exp\left(-\frac{1}{2}\chi^2({\bf p})\right) =
 \exp\left[-\frac{1}{2} \sum_{n=1}^{N}
 \frac{\Big(\mc{O}_{n}^{\rm (th)}({\bf p})-
               \mc{O}_{n}^{\rm (exp)}\Big)^{2}}
 {\Delta\mc{O}_{n}^{2}}\right] \;.
 \label{Likelihood}
\end{equation}
Although often regarded as a probability distribution, the absolute
normalization of the likelihood function is not required. Rather,
the merit of the likelihood function lies in its \emph{relative}
value. This is sufficient to decide which of two parameter sets (say
$\psub{1}$ and $\psub{2}$) is more likely to reproduce the
experimental data---or ultimately to obtain the most likely
parameter set $\psub{0}$. Note that the most likely, or
\emph{optimal}, parameter set $\psub0$ is the one that minimizes the
negative of the ``log-likelihood" function which is, up to a
constant, the objective $\chi^{2}$-function:
\begin{equation}
 -\ln{\mc L}({\bf p}) =
  \frac{1}{2}\chi^2({\bf p}) = \frac{1}{2} \sum_{n=1}^{N}
  \frac{\Big(\mc{O}_{n}^{\rm (th)}({\bf p})-
               \mc{O}_{n}^{\rm (exp)}\Big)^{2}}
               {\Delta\mc{O}_{n}^{2}} \;.
 \label{logLikelihood}
\end{equation}
That is, the optimal parameter set satisfies
\begin{equation}
 \frac{\partial\chi^{2}({\bf p})}{\partial p_{i}}
 \Big|_{{\bf p}={\bf p}_{0}} \equiv
 \partial_{i}\chi^{2}(\psub{0})\!=\!0 \quad
 ({\rm for\;} i=1,\ldots,F)\;.
 \label{chi2min}
\end{equation}
The existence of a minimum---as opposed to a maximum or saddle
point---also implies that a particular set of $F$ second derivatives
must all be positive. As we will argue shortly, the $F\!\times\!F$
matrix of second derivatives evaluated at the minimum $\psub{0}$
contains a wealth of information that is often lost from traditional
approaches that are limited to the extraction of the optimal parameter
set. To uncover such a wealth of information it is both illuminating
and highly intuitive to exploit the probabilistic nature of the
likelihood function. In particular, one can efficiently sample the
parameter space via a standard Metropolis Monte-Carlo
algorithm. Indeed, at the end of such process, one would have
generated a ``Markov chain" of models $\{{\bf p}_{1},{\bf
p}_{2},\ldots\,{\bf p}_{M}\}$ that approaches the desired likelihood
function in the limit of $M\!\gg\!1$. Moreover, as one generates the
distribution of models---and hence calculates all $N$ observables
appearing in the objective function for each member of the Markov
chain---one can easily generate the full probability
distribution---not only the average---for each observable.  Denoting
by $\mc{A}$ such an observable---or indeed any generic
observable---one defines the probability $P(\mc{A})$ as follows:
\begin{equation}
   P(\mc{A}) = \frac{\int d{\bf p}\,
   \delta\!\left(\mc{A}\!-\!\mc{A}^{\rm (th)}({\bf p})\right)
   \exp\left(-\frac{1}{2}\chi^2({\bf p})\right)} {\int d{\bf p}
   \exp\left(-\frac{1}{2}\chi^2({\bf p})\right)} \,,
 \label{PofA}
\end{equation}
where $\mc{A}^{\rm (th)}({\bf p})$ denotes the theoretical prediction
by model ${\bf p}$ of observable $\mc{A}$. In particular, Monte-Carlo
evaluation of the integral provides a rather simple and convenient
description of the probability distribution in histogram form. That
is, we define the normalized probability that the observable $\mc{A}$
may be found in an interval centered at $\mc{A}_{n}$ of width $\Delta$
as
\begin{equation}
   P_{\!\Delta}(\mc{A}_{n}) \equiv
   \int_{\mc{A}_{n}\!-\!\Delta/2}^{\mc{A}_{n}\!+\!\Delta/2}
    P(\mc{A}) d\mc{A} =
    \frac{\int d{\bf p}\,
    \delta_{\!\Delta}\!\left(\mc{A}_{n}\!-\!\mc{A}^{\rm (th)}({\bf p})\right)
   \exp\left(-\frac{1}{2}\chi^2({\bf p})\right)} {\int d{\bf p}
   \exp\left(-\frac{1}{2}\chi^2({\bf p})\right)} =
   \frac{1}{M}\!\sum_{m=1}^{M}
   \delta_{\!\Delta}\!\Big(\mc{A}_{n}\!-\!\mc{A}^{\rm (th)}_{m}\Big) \;,
  \label{PofAn}
\end{equation}
where $\mc{A}^{\rm (th)}_{m}\!\equiv\!\mc{A}^{\rm (th)}({\bf p}_{m})$
and $\delta_{\!\Delta}(x\!-\!y)$ is a \emph{smeared} Dirac delta function
of width $\Delta$ defined as follows:
\begin{equation}
 \delta_{\!\Delta}(x\!-\!y) = \int_{y-\Delta/2}^{y+\Delta/2}
 \delta(x\!-\!y')\,dy'=
  \left\{ \begin{array}{ll}
  1 & \mbox{if $|x-y|\!\le\!\Delta/2$\,;} \\
  0 & \mbox{otherwise\,.}
 \end{array} \right.
 \label{SmDelta}
\end{equation}
Note that the operational evaluation of the last term in
Eq.\,(\ref{PofAn}) is straightforward. Indeed, for each value
$\mc{A}^{\rm (th)}_{m}$ of the total of $M$ values generated in the
Monte-Carlo simulation, one simply increases by one unit the bin
that houses such value. That is, one increases the bin number $n$
that satisfies $\mc{A}_{n}\!-\!\Delta/2\!\le\!\mc{A}^{\rm (th)}_{m}
\!<\!\mc{A}_{n}\!+\!\Delta/2$. The histogram generated in this
manner provides a faithful representation of the continuous
probability distribution $P(\mc{A})$, at least in the limit of
$\Delta\!\ll\!1$ and $M\!\gg\!1$.  In particular, the average value
of $\mc{A}$ is simply given by
\begin{equation}
 \langle \mc{A} \rangle =
  \int_{-\infty}^{\infty} \mc{A} P(\mc{A}) d\mc{A} =
  \frac{1}{M}\!\sum_{m=1}^{M} \mc{A}^{\rm (th)}_{m} \,.
 \label{AvgA}
\end{equation}
Moreover, given that the full probability distribution $P(\mc{A})$
has been generated, one is now in a position to evaluate any
of its moments. In particular, a moment---or rather a combination of
moments---of great significance is the variance of the distribution.
That is,
\begin{equation}
 \sigmasub{\mc{A}}^{2} =
 \langle \mc{A}^{2} \rangle \!-\!
  \langle \mc{A} \rangle^{2} \quad {\rm with} \quad
   \langle \mc{A}^{2} \rangle =
   \frac{1}{M}\!\sum_{m=1}^{M} \left(\mc{A}^{\rm (th)}_{m}\right)^{2} \,.
  \label{VarA}
\end{equation}
The variance is enormously insightful as it represents a proper
statistical measure of the \emph{theoretical} uncertainty in the
determination of the average value of the observable. Indeed,
knowledge of the average value---by itself---is of limited utility
as it provides no information on whether the associated probability
distribution is narrow or broad. If broad, it suggests that the
observables adopted in the calibration protocol are weakly
correlated to $\mc{A}$. A notable example of such a situation is
$L$, the slope of the symmetry energy at saturation density. Indeed,
nuclear binding energies and charge radii that have been
traditionally used in the calibration of the EDF are largely
insensitive to the density dependence of the symmetry energy. This
information is highly valuable as it suggests a clear path for
improving the EDF. In this context, introducing the concept of
covariance between two observables $\mc{A}$ and $\mc{B}$ is
appropriate. That is,
\begin{equation}
 {\rm cov}(\mc{A},\mc{B}) \equiv
  \langle\mc{A}\mc{B}\rangle \!-\!
  \langle\mc{A}\rangle\langle\mc{B}\rangle \,,
 \label{CovAB}
\end{equation}
where
\begin{equation}
   \langle\mc{A}\mc{B}\rangle =
   \int_{-\infty}^{\infty} d\mc{A}
   \int_{-\infty}^{\infty} d\mc{B}\,
   \mc{A}\mc{B}P(\mc{A},\mc{B}) =
   \frac{1}{M}\!\sum_{m=1}^{M}
   \left(\mc{A}^{\rm (th)}_{m}\mc{B}^{\rm (th)}_{m}\right)\,.
  \label{BraABKet}
\end{equation}
Finally, the \emph{Pearson product moment correlation coefficient},
or simply the correlation coefficient,
$\large{\rhosub{$\!\!\mc{AB}$}}$ is defined as follows:
\begin{equation}
 {\large{\rhosub{$\!\!\mc{AB}$}}} \equiv
 \frac{{\rm cov}(\mc{A},\mc{B})}
 {\sigmasub{\mc{A}}\sigmasub{\;\mc{B}}} \,.
 \label{CorrAB}
\end{equation}
The correlation coefficient ${\large{\rhosub{$\!\!\mc{AB}$}}}$ has an intuitive
geometric interpretation. By defining the following two unit vectors in
$M$-dimensions as
\begin{equation}
 {a}_{m} \equiv \frac{1}{\sqrt{M}}
 \left(\frac{\mc{A}_{m}-\langle\mc{A}\rangle}{\sigmasub{\mc{A}}}\right)
 \quad{\rm and}\quad
 {b}_{m} \equiv \frac{1}{\sqrt{M}}
 \left(\frac{\mc{B}_{m}-\langle\mc{B}\rangle}{\sigmasub{\;\mc{B}}}\right)\,,
 \label{CosAB}
\end{equation}
the correlation coefficient can be written as the cosine of the angle
between these two unit vectors. That is,
\begin{equation}
  {\rhosub{$\!\!\mc{AB}$}}= \hat{a}\cdot\hat{b}
  \equiv\cos(\hat{a},\hat{b}) \;.
 \label{Corrab}
\end{equation}
A value of ${\large{\rhosub{$\!\!\mc{AB}$}}}\!=\!\pm 1$ means that the
two observables are fully correlated/anti-correlated, whereas a value
of ${\large{\rhosub{$\!\!\mc{AB}$}}}\!=\!0$ means that the observables
are totally uncorrelated.

\subsection{Gaussian Approximation to the Likelihood Function}
\label{Gaussian}

Whereas the statistical treatment based on the likelihood function
is free of approximations, a highly insightful analytic approach
relies on a gaussian approximation to the likelihood function. In
this approach one limits the exploration of the landscape to the
immediate vicinity of the $\chi^{2}$ minimum (or maximum
likelihood). Denoting by $\psub0$ the \emph{optimal} parameter set,
the gaussian approximation consists of studying the small
(quadratic) oscillations around the $\chi^{2}$ minimum. That is,
\begin{equation}
 \chi^2({\bf p}) \approx \chi^{2}(\psub0) + \frac{1}{2}\sum_{i,j=1}^{F}
 ({\bf p}-\psub0)_{i} ({\bf p}-\psub0)_{j}
 \partial_{i}\partial_{j}\chi^{2}(\psub0) \,.
 \label{Taylor1}
\end{equation}
Having found the absolute minimum $\psub0$---undoubtedly
one of the hardest numerical task---it is now convenient to 
quantify the deviations from the minimum in terms of the 
following scaled variables:
\begin{equation}
  x_{i} \equiv \frac{({\bf p}-\psub0)_{i}}{(\psub0)_{i}} \;.
 \label{xDef}
\end{equation}
In terms of these scaled variables, Eq.(\ref{Taylor1}) takes the
following compact form:
\begin{equation}
 \chi^2({\bf p}) = \chi^{2}_{0} +
 {\bf x}^{T}{\hat{\mc M}}\,{\bf x} \;,
 \label{Taylor2}
\end{equation}
where ${\bf x}$ is a column vector of dimension $F$, ${\bf x}^{T}$
is the corresponding transpose (row) vector, and ${\hat{\mc M}}$
is the {\sl symmetric} $F\!\times\!F$ matrix of second derivatives.
That is,
\begin{equation}
 {\mc M}_{ij} = \frac{1}{2}\left(\frac{\partial^2\chi^{2}}
 {\partial x_{i}\partial x_{j}}\right)_{{\bf x}=0} =
 \frac{1}{2}\partial_{i}\partial_{j}\chi^{2}_{0}\,.
 \label{MMatrix}
\end{equation}

In the gaussian approximation the probability distribution of any generic
observable $\mc{A}$ may now be evaluated in closed form. Starting
from Eq.\,(\ref{PofA}) one obtains in the gaussian approximation
\begin{equation}
   P(\mc{A}) = \frac{\int d{\bf x}\,
   \delta\!\left(\mc{A}\!-\!\mc{A}_{0}-x_{i}\mc{A}_{i}\right)
   \exp\left(-\frac{1}{2}{\bf x}^{T}{\hat{\mc M}}\,{\bf x}\right)} {\int d{\bf x}
   \exp\left(-\frac{1}{2}{\bf x}^{T}{\hat{\mc M}}\,{\bf x}\right)} =
   \int_{-\infty}^{\infty}\,\frac{dp}{2\pi} \left[\frac{Z({\bf J})}{Z(0)}\right]
   \exp\!\Big(ip(\mc{A}\!-\!\mc{A}_{0})\Big) \,,
 \label{PofA1}
\end{equation}
where $\mc{A}_{0}\!\equiv\!\mc{A}^{\rm (th)}_{0}$,
$\mc{A}_{i}\!\equiv\!\partial_{i}\mc{A}^{\rm (th)}_{0}$,
and $J_{i}\!\equiv\!-ip\mc{A}_{i}$. Note that we have introduced
the {\sl ``partition''} function as
\begin{equation}
  Z({\bf J}) = \int d{\bf x}\, \exp\left(-\frac{1}{2}
  {\bf x}^{T}{\hat{\mc M}}\,{\bf x}+
  {\bf J}\cdot{\bf x}\right)  =
  Z(0)\exp\left(\frac{1}{2}
  {\bf J}^{T}{\hat{\mc M}^{-1}}\,{\bf J}\right) \,,
\label{PartitionFcn}
\end{equation}
where the gaussian integral was evaluated by completing the square. The
remaining integral in $p$---also gaussian---can be performed by
elementary means and this yields the following results for $P(\mc{A})$
in the gaussian approximation:
\begin{equation}
   P(\mc{A}) =
   \frac{1}{\sqrt{2\pi{\sigmasub{\!\mc{A}}^{2}}}}
   \exp\!\left(-\frac{(\mc{A}\!-\!\mc{A}_{0})^{2}}
                             {2{\sigmasub{\!\mc{A}}^{2}}}\right) \,,
 \label{PofA2}
\end{equation}
where the variance of $\mc{A}$ is given by
\begin{equation}
 {\sigmasub{\!\mc{A}}^{2}} =
  {\bf A}^{\!T} \hat{{\mc M}}^{-1} {\bf A} \equiv
 \sum_{i,j=1}^{F}
 \left(\frac{\partial\mc{A}}{\partial x_{i}}\right)_{\!0}
  {\mc M}^{-1}_{ij}
 \left(\frac{\partial\mc{A}}{\partial x_{j}}\right)_{\!0} \,.
\label{VarianceA}
\end{equation}
Another useful and widely used statistical concept is that of a
\emph{confidence interval}, which represents the probability that the
``true" value of an observable falls within a given interval. In the
case of an observable $\mc{A}$ with a gaussian distribution as in
Eq.\,(\ref{PofA2}) it is given by
\begin{equation}
 P(n) \equiv \int_{\mc{A}_{0}-n\sigmasub{\mc{A}}}^{\mc{A}_{0}+n\sigmasub{\mc{A}}}
 P(\mc{A})d\mc{A} =
 \int_{\mc{A}_{0}-n\sigmasub{\mc{A}}}^{\mc{A}_{0}+n\sigmasub{\mc{A}}}
 \frac{d\mc{A}}{\sqrt{2\pi\sigmasub{\!\mc{A}}^{2}}}
 \exp\left(-\frac{(\mc{A}-\mc{A}_{0})^{2}}{2\sigmasub{\!\mc{A}}^{2}}\right)=
 {\rm erf}\left(\frac{n}{\sqrt{2}}\right) =
  \begin{cases}
    0.683 & \text{if } n=1 \\
    0.955 & \text{if } n=2 \\
    0.997 & \text{if } n=3
  \end{cases}\,,
\end{equation}
where ${\rm erf}(x)$ is the ``error'' function. This result implies
that there is a better than 99\% probability that the actual value of
the observable $\mc{A}$ is contained in an interval within $\pm 3$
standard deviations from its average value. Equivalently, it is
expected that 99.7\% of the points generated, for example via a
Monte-Carlo simulation, will fall within such interval---at least in
the limit in which the exact probability distribution is closely
approximated by a gaussian.

Although it is well known that the variance reflects the spread of an
observable around its average value, analyzing the sources driving
such a spread in Eq.\,(\ref{VarianceA}) is illuminating. To do so, we
first bring the symmetric matrix of second derivatives into a diagonal
form by means of an orthogonal transformation. That is,
$\hat{{\mc M}}\!=\!\hat{O}{\hat{\mc D}}{\hat{O}^{T}}$, where $\hat{O}$
is the matrix of eigenvectors and $\hat{\mc D}$ is the \emph{diagonal}
matrix of eigenvalues
${\hat{\mc D}}\!=\!{\rm diag}(\lambda_{1},\ldots,\lambda_{F})$.
Note that all the eigenvalues are positive by virtue that the
objective function attains its minimum value at ${\bf x}\!=\!0$. In
terms of the positive eigenvalues of the matrix of second derivatives
one can then write
\begin{equation}
 {\sigmasub{\mc{A}}^{2}} =
  {\bf A}^{\!T}\Big(\hat{O}{\hat{\mc D}^{-1}}{\hat{O}^{T}}\Big){\bf A} =
  (\hat{O}^{T}{\bf A})^{T}\hat{\mc D}^{-1}(\hat{O}^{T}{\bf A})=
  \sum_{i=1}^{F} \lambda_{i}^{-1}
  \left(\frac{\partial\mc{A}}{\partial
  \xi_{i}}\right)^{\!2}_{\!0}\,,
\label{VarianceA1}
\end{equation}
where the $F$-dimensional vector ${\bm\xi}\!=\!\hat{O}^{T}{\bf x}$
represents a point in parameter space expressed, not in terms of the
original model parameters but rather, in terms of a linear
combination of them ({\sl i.e.,} in terms of the new ``rotated''
basis).  In particular, each eigenvalue $\lambda_{i}$ controls the
deterioration in the quality of the fit as one moves along a
direction defined by its corresponding eigenvector. A ``soft'' or
``flat" direction---characterized by a small or zero eigenvalue 
$\lambda_{i}$ and thus little deterioration in the quality 
measure---involves a particular linear combination of model 
parameters that is poorly
constrained by the choice of observables included in the definition
of the objective function\,\cite{Bertsch:2004us}.  Moreover, if the
observable of interest $\mc{A}$ is sensitive to such a direction, in
the sense that the magnitude of its derivative
$\partial\mc{A}/\partial \xi_{i}$ is large, then the overall
variance ${\sigmasub{\mc{A}}^{2}}$ will also be large. However, by
properly identifying such a soft direction (or directions) one can
readily propose improvements to the model. Again, a relevant example
of great theoretical and experimental interest involves the slope of
the symmetry energy $L$. Although not itself an experimental
observable, $L$ is of critical importance to the equation of state
of neutron-star matter and ultimately to such astrophysical
observables as the neutron-star radius. Yet predictions for $L$ are
usually accompanied by large theoretical errors, suggesting that the
observables incorporated in the objective function (such as masses
and charge radii) are largely insensitive to the combination of
isovector parameters that constrain the behavior of $L$. However,
one has been able to identify experimental observables---notably the
neutron-skin thickness and electric dipole polarizability of heavy
nuclei---that are strongly correlated to $L$. Ongoing experimental
programs at various facilities around the world are actively engaged
in measuring these isovector observables that will provide vital
constraints on the density dependence of the symmetry energy.

Given the critical role of correlations between observables, we
conclude this section by computing the joint probability distribution
of two generic observables $\mc{A}$ and $\mc{B}$ in the gaussian
approximation. That is,
\begin{align}
   P(\mc{A},\mc{B}) &= \frac{\int d{\bf x}\,
   \delta\!\left(\mc{A}\!-\!\mc{A}_{0}-x_{i}\partial_{i}\mc{A}\right)
   \delta\!\left(\mc{B}\!-\!\mc{B}_{0}-x_{i}\partial_{i}\mc{B}\right)
   \exp\left(-\frac{1}{2}{\bf x}^{T}{\hat{\mc M}}\,{\bf x}\right)} {\int d{\bf x}
   \exp\left(-\frac{1}{2}{\bf x}^{T}{\hat{\mc M}}\,{\bf x}\right)} \nonumber \\
   &=
   \int_{-\infty}^{\infty}\,\frac{dp_{a}}{2\pi}
   \int_{-\infty}^{\infty}\,\frac{dp_{b}}{2\pi}
   \left[\frac{Z({\bf J})}{Z(0)}\right]
   \exp\!\Big(ip_{a}(\mc{A}\!-\!\mc{A}_{0})+ip_{b}(\mc{B}\!-\!\mc{B}_{0})\Big) \,,
 \label{PofAB}
\end{align}
where $Z({\bf J})$ is identical to expression given in
Eq.\,(\ref{PartitionFcn}) but with ${\bf J}$ now being defined as
$J_{i}\!\equiv\!-i(p_{a}\partial_{i}\mc{A}+p_{b}\partial_{i}\mc{B})$.
The remaining two-dimensional integral over $p_{a}$ and $p_{b}$ is
also a gaussian integral that may be computed by completing the
square.  One obtains,
\begin{equation}
   P(\mc{A},\mc{B}) =
   \frac{\exp\!\left(-\frac{1}{2}j^{T}\hat{\Sigma}^{-1}j\right)}
   {2\pi\sqrt{\det{\hat{\Sigma}}}}\,
 \label{PofAB2}
\end{equation}
where $j^{T}\!\equiv\!(\mc{A}\!-\!\mc{A}_{0},\mc{B}\!-\!\mc{B}_{0})$ and
$\hat{\Sigma}$ is the symmetric $2\!\times\!2$ \emph{covariance} matrix
defined as follows:
\begin{equation}
   \hat{\Sigma} =
    \begin{pmatrix}
      {\rm cov}(\mc{A},\mc{A}) & {\rm cov}(\mc{A},\mc{B}) \\
      {\rm cov}(\mc{B},\mc{A}) & {\rm cov}(\mc{B},\mc{B})
    \end{pmatrix} =
    \begin{pmatrix}
       \sigmasub{\mc{A}}^{2} & \sigmasub{\mc{A}}\sigmasub{\,\mc{B}}
       \rhosub{$\!\!\mc{AB}$} \\
       \sigmasub{\mc{A}}\sigmasub{\,\mc{B}}\rhosub{$\!\!\mc{AB}$} &
       \sigmasub{\,\mc{B}}^{2}
    \end{pmatrix} \;.
 \label{CovMtrx}
\end{equation}
Note that the covariance matrix contains information about the spread
of the individual observables plus the correlation among them. Moreover,
in the gaussian approximation the covariance of $\mc{A}$ and $\mc{B}$
is given as the natural extension of Eq.\,(\ref{VarianceA}). That is,
\begin{equation}
 {\rm cov}(\mc{A},\mc{B}) =
 {\bf A}^{\!T} \hat{{\mc M}}^{-1} {\bf B} =
 \sum_{i,j=1}^{F}
 \left(\frac{\partial\mc{A}}{\partial x_{i}}\right)_{\!0}
   {\mc M}^{-1}_{ij}
 \left(\frac{\partial\mc{B}}{\partial x_{j}}\right)_{\!0} \,.
\label{CovAB1}
\end{equation}

We conclude this section with a discussion of the two-dimensional
generalization of the confidence interval, namely, the
\emph{confidence ellipse}. As in the case of the confidence
interval, we want to quantify the likelihood that a pair of
observables fall within a given two-dimensional region. The notion
of a confidence ellipse is motivated by the exponent given in
Eq.\,(\ref{PofAB2}). Defining $a\!\equiv\mc{A}\!-\!\mc{A}_{0}$ and
$b\!\equiv\!\mc{B}\!-\!\mc{B}_{0}$, such an exponent may be written
as follows:
\begin{equation}
 j^{T}\hat{\Sigma}^{-1}j = \psi^{T}\hat{\Sigma}_{D}^{-1}\psi=
 \frac{{x}^{2}}{\sigma_{\!x}^{2}}+ \frac{{y}^{2}}{\sigma_{\!y}^{2}}\equiv
 R^{2} \;,
 \label{CovEllipse}
 \end{equation}
where we have used the fact that the symmetric $2\!\times\!2$
covariance matrix $\hat{\Sigma}$ can be brought into diagonal form by
means of an orthogonal transformation. Here $\sigma_{\!x}^{2}$ and
$\sigma_{\!y}^{2}$ are the two positive eigenvalues of $\hat{\Sigma}$
and $\psi^{T}\!\equiv\!(x,y)$ represents the coordinates of the
two-dimensional point in the rotated coordinate system. In this
rotated (and shifted) coordinate system, Eq.\,(\ref{CovEllipse})
represents the equation of an ellipse centered at the origin with
semi-major axis $R\sigma_{x}$ and semi-minor axis $R\sigma_{y}$
(assuming $\sigma_{\!x}\!\ge\!\sigma_{\!y}$). The confidence ellipse
of ``magnitude" $R$ represents the probability that the ``true" values
of both observables lie within such an ellipse. Such a probability is
given by
\begin{equation}
 P(R) \equiv \int_{\Omega(R)} P(\mc{A},\mc{B})\,d{\mc{A}}\,d{\mc{B}}=
  \int_{\Omega(R)}
  \frac{\exp\!\left(-\frac{1}{2}j^{T}\hat{\Sigma}^{-1}j\right)}
  {2\pi\sqrt{\det{\hat{\Sigma}}}}\,d{\mc{A}}\,d{\mc{B}}=
  \Big(1\!-\!e^{-R^{2}/2}\Big) =
  \begin{cases}
    0.3935 & \text{if } R=1 \\
    0.9500 & \text{if } R=2.4478 \\
    0.9900 & \text{if } R=3.0349
  \end{cases}\,,
\end{equation}
where $\Omega(R)$ represents the elliptical region defined by
Eq.\,(\ref{CovEllipse}). In particular, this implies that 95\% of the
points generated, for example via a Monte-Carlo simulation, will lie
within the $R\!=\!2.4478$ confidence ellipse---provided the actual
probability distribution is well approximated by Eq.\,(\ref{PofAB2}).

\section{Results}
\label{Results} We devote this section to illustrate some of the
formal ideas developed earlier. Our aim is to do so by using a few
simple examples that demonstrate the power and elegance of the
approach. The first example allows for an exact implementation of
the maximum likelihood method by focusing on a simple exercise with
very modest computational demands: \emph{the liquid drop model}. In
the second example we discuss the calibration of a relativistic EDF
where the computational demands are severe enough to limit the
approach to the gaussian approximation. Finally, we briefly address
the role of systematic errors by focusing on the correlation between
the neutron-skin thickness of ${}^{208}$Pb and the slope of the
symmetry energy $L$.

\subsection{Liquid Drop Model: Calibration and Correlations}
\label{LDM}

The semi-empirical mass formula of Bethe and
Weizs\"acker---conceived shortly after the discovery of the neutron
by Chadwick---treats the nucleus as an incompressible liquid drop
consisting of two quantum fluids: one neutral and one charged. For a
liquid drop consisting of $Z$ protons, $N$ neutrons, and a total
baryon number $A\!=\!Z\!+\!N$, the mass formula may be written in
general in terms of the individual nucleon masses ($m_{p}$ and
$m_{n}$) and the nuclear binding energy $B(Z,N)$ that contains all
the complicated nuclear dynamics:
$M(Z,N)\!=\!Zm_{p}\!+\!Nm_{n}\!-\!B(Z,N)$. In the particular context
of the liquid drop model (LDM), the binding energy is written in
terms of a handful of empirical parameters that encapsulate the
underlying physics of a charged quantum drop. That is,
\begin{equation}
 B(N,Z)\!=\!a_{\rm v}A - a_{\rm s}A^{2/3} -
 a_{\rm c}\!\frac{Z^{2}}{A^{1/3}} - a_{\rm a}\!\frac{(N\!-\!Z)^{2}}{A}+\ldots
 \label{BWMF}
\end{equation}
Note that as defined in Eq.\,(\ref{BWMF}), all four empirical
parameters are positive definite. Also note that in general the LDM
involves additional parameters, such as terms associated with pairing
correlations and with a surface symmetry energy, but for our purposes
these four are sufficient.  The volume term ($a_{\rm v}$) represents
the binding energy of a macroscopic and symmetric
($Z\!=\!N\!=\!A/2\!\gg\!1$) drop in the absence of Coulomb forces. In
turn, the next three terms represent repulsive terms associated with
the development of a nuclear surface ($a_{\rm s}$), the Coulomb
repulsion among protons ($a_{\rm c}$), and the Pauli exclusion
principle that favors symmetric systems ($a_{\rm a}$).

The aim of the present exercise is to obtain the optimal set of
empirical LDM parameters that best describes the binding energies of
all 576 even-even nuclei with $N\!\ge\!8$ and $Z\!\ge\!8$, as
tabulated in the AME-2003 atomic mass evaluation of Audi, Wapstra, and
Thibault\,\cite{Audi:2002rp}. To this end we define the following
objective function:
\begin{equation}
 \chi^{2}(a_{\rm v},a_{\rm s},a_{\rm c},a_{\rm a})=
\ \sum_{n=1}^{576}
  \frac{\Big[B^{\rm (th)}(N_{n},Z_{n})-
                B^{\rm (exp)}(N_{n},Z_{n})\Big]^{2}}
               {\Delta B(N_{n},Z_{n})^{2}} \;,
 \label{Chi2LDM}
\end{equation}
where the sum is over all 576 even-even nuclei, $B^{\rm (th)}(N,Z)$ is
the LDM prediction, and $B^{\rm (exp)}(N,Z)$ is the quoted AME-2003
value for the binding energy\,\cite{Audi:2002rp}. Given that most
experimental masses have been measured with enormous precision, we
adopt the same \emph{theoretical} error as in
Ref.\,\cite{Dobaczewski:2014jga} of this focus issue. That is,
\begin{equation}
 \Delta B(N_{n},Z_{n})\!=\!\Delta B^{\rm {(th)}}\!=\!3.8\,{\rm MeV}.
 \label{errorLDM}
\end{equation}
This choice of theoretical error will allow for a meaningful
comparison against the results obtained in
Ref.\,\cite{Dobaczewski:2014jga} by different means. In our
case---given that the computational demands involved in the evaluation
of the LDM objective function are so modest---we have opted to
generate the full probability distribution associated with the
likelihood function without relying on the gaussian
approximation. That is, we have sampled the four-dimensional LDM
parameter space by employing the following likelihood function:
\begin{equation}
 {\mc L}\Big(a_{\rm v},a_{\rm s},a_{\rm c},a_{\rm a}
 \Big|B_{n}^{\rm (exp)}\Big) =
 \exp\left[-\frac{1}{2}\chi^{2}(a_{\rm v},a_{\rm s},a_{\rm c},a_{\rm a})\right]\,.
 \label{LFLDM}
\end{equation}
The sampling of the parameter space was done via a standard
Metropolis-Monte-Carlo algorithm\,\cite{Metropolis:1953am}. Initially,
a Markov chain consisting of 2 million configurations was generated in
an effort to thermalize the system. After the system has been properly
thermalized, another set of 2 million configurations was
generated---with 1 of every 10 configurations kept to perform the
statistical analysis.  In this manner, a total of $M\!=\!200,000$
``quartets" of LDM parameters $\{(a_{\rm v}^{m},a_{\rm s}^{m},a_{\rm
c}^{m},a_{\rm a}^{m}), m=1,M\}$ was generated according to the
likelihood function. Using such a data set, a standard statistical
analysis was performed.

\begin{figure}[ht]
\vspace{-0.05in}
\includegraphics[width=0.6\columnwidth,angle=0]{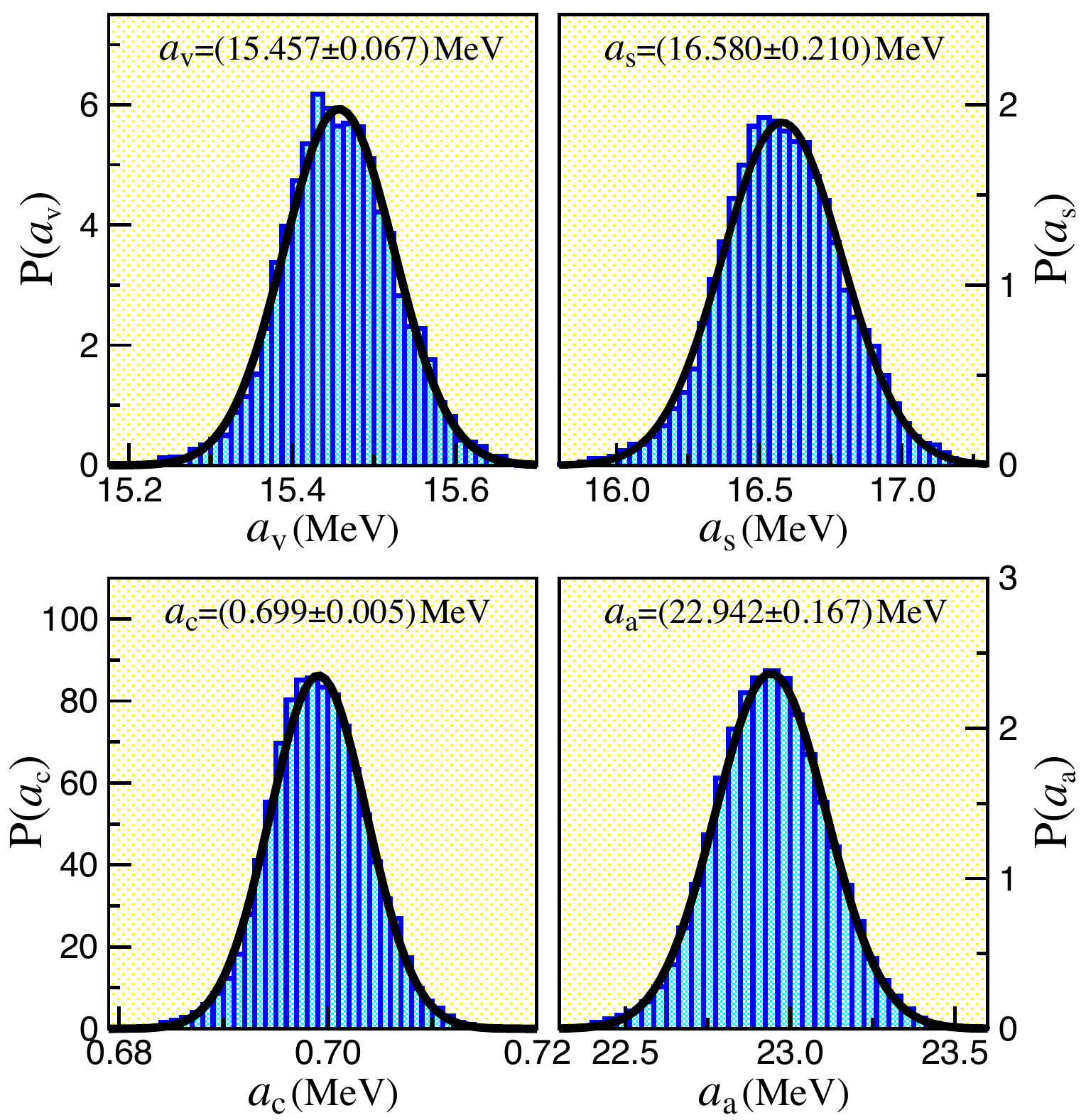}
\caption{(color online) Probability distribution for the four empirical
parameters of the liquid-drop formula as obtained via a
Metropolis-Monte-Carlo method. The black solid line and the associated
labels represent the results obtained in the gaussian approximation as
given in Eq.\,(\ref{PofA2}).}
\label{Fig1}
\end{figure}

In Fig.\ref{Fig1} we display (in histogram form) the probability
distribution for each of the four empirical LDM parameters. Each panel
also lists the average and standard deviation of the ``raw"
Monte-Carlo data. Using these parameters, gaussian approximations to
the normalized histograms were generated and are displayed in the
figure with the continuous (black) line. In all four cases the
gaussian approximation provides a faithful representation of the
Monte-Carlo data. Although obtained by different means, the extracted
averages and errors are all fully consistent with those reported in
Table\,1 of Ref.\,\cite{Dobaczewski:2014jga}. Note that without
properly reported errors, it is difficult to assess whether the
experimental input used to define the objective function is rich
enough to constrain all model parameters. With errors of the order of
1\%, we can conclude that the 576 experimental masses are sufficient
to properly constrain all four LDM parameters.

\begin{figure}[ht]
\vspace{-0.05in}
\includegraphics[width=2.75in,height=3in,angle=0]{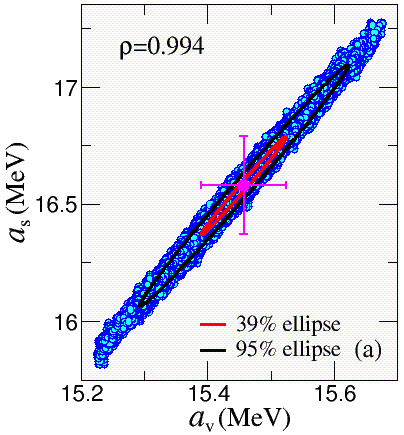}
 \hspace{0.25cm}
\includegraphics[width=2.75in,height=3in,angle=0]{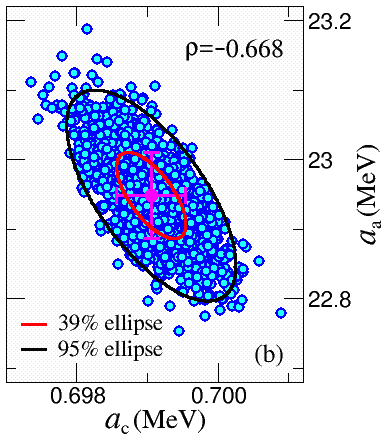}
\caption{(color online) (a) Correlation plot between the volume and
surface terms of the liquid-drop model. The 200,000 points represent
the raw data obtained from the Monte-Carlo simulation. Also shown are
the 39\% (in red) and 95\% (in black) confidence ellipses. (b) Same
as in (a) but now for the Coulomb and asymmetry terms of the LDM.
Here the 1,784 Monte-Carlo points were obtained by performing
suitable selection cuts as described in the text.}
\label{Fig2}
\end{figure}

Whereas extracting averages and errors is highly informative, the
greatest power of the statistical analysis lies in assessing
correlations between observables. As stated earlier, quantities that
may be difficult---or impossible---to measure, may be significantly
constrained by finding an experimental observable that is strongly
correlated to it. We have used the Monte-Carlo data to compute
correlation coefficients between the various LDM parameters and have
listed their values in Table\,\ref{Table1}.
\begin{table}[h]
  \begin{tabular}{|c||c|c|c|c|}
   \hline
    $\rhosub{AB}$ & $a_{\rm v}$ & $a_{\rm s}$ & $a_{\rm c}$   & $a_{\rm a}$ \\
   \hline
   \hline
    $a_{\rm v}$  & 1.000 & 0.994 & 0.986 & 0.920 \\
    $a_{\rm s}$  & 0.994 & 1.000 & 0.965 & 0.907 \\
    $a_{\rm c}$  & 0.986 & 0.965 & 1.000 & 0.888 \\
    $a_{\rm a}$  & 0.920 & 0.907 & 0.888 & 1.000 \\
   \hline
  \end{tabular}
  \caption{Correlation coefficients between the four LDM parameters.}
  \label{Table1}
 \end{table}
In all cases the correlation coefficients are fairly large,
indicating that increasing the value of the nuclear attraction
$a_{\rm v}$ must be properly compensated by a corresponding increase
in the remaining three repulsive parameters. Pictorially, such
correlations may be displayed by plotting the Monte-Carlo data for
any given pair of quantities. In particular, in Fig.\,\ref{Fig2}a we
display the correlation between the volume ($a_{\rm v}$) and surface
($a_{\rm s}$) terms in the liquid-drop formula.  Averages and errors
for these two quantities are indicated by the central cross.
Moreover, we display confidence ellipses computed as described
earlier in the formalism section. The smaller of the two (displayed
in red) represents the 39\% confidence ellipse whereas the larger
one (displayed in black) depicts the 95\% confidence region. That
is, in the limit in which the exact probability distribution may be
accurately approximated by a gaussian distribution, only 5\% (or
10,000) of a total of 200,000 Monte-Carlo points fall outside the
95\% confidence ellipse.  Note the very large eccentricity of the
confidence ellipses. This is because in the limit in which the
correlation coefficient approaches $\pm 1$, the confidence ellipse
``degenerates'' into a straight line, as the ratio of the semi-minor
to semi-major axes goes to zero.

Despite the wealth of information contained in the correlation
plots, one should exercise caution in interpreting the results. For
example, Table\,\ref{Table1} reveals a large and \emph{positive}
correlation coefficient (of about $+0.9$) between the Coulomb and
asymmetry terms. At first sight this result may seem
counter-intuitive; shouldn't these two quantities be
\emph{anti-}correlated?  That is, shouldn't the asymmetry term
\emph{decrease} if the Coulomb repulsion \emph{increases} in order
to maintain the LDM predictions close to the experimental values?
The answer to this apparent contradiction lies in the fact that in
generating the correct distribution of LDM parameters, all four
parameters become inextricably linked. And the most ``efficient''
response to an increase in the value of the dominant volume term is
for the three remaining parameters to all increase accordingly.
Hence, to reveal the expected anti-correlation between $a_{\rm c}$
and $a_{\rm a}$ one should provide selection cuts---or ``gates"---to
filter data in which the other parameters, in this case $a_{\rm v}$
and $a_{\rm s}$, are nearly fixed. To do so---and still have enough
data to perform the correlation analysis---we have only selected
those Monte-Carlo points in which both $a_{\rm v}$ and $a_{\rm s}$
are within 0.04\,$\sigma$ of their average values.  The resulting
1,784 points (out of 200,000) are plotted in Fig.\,\ref{Fig2}b
alongside the 39\% and 95\% confidence ellipses. It is now evident
that $a_{\rm c}$ and $a_{\rm a}$ are indeed anti-correlated. That
is, if both $a_{\rm v}$ and $a_{\rm s}$ are (nearly) fixed at their
average values, then an increase in the Coulomb repulsion must be
accompanied by a corresponding decrease in the asymmetry term in
order to provide the best description of the experimental data.

\subsection{Relativistic Energy Density Functional: Calibration and Correlations}
\label{RDF}

In this section we implement the statistical formulation developed
earlier to the calibration of a relativistic EDF. In contrast to the
liquid-drop model, the high computational demands of the present
case preclude us from generating the exact probability distribution.
Thus, all our results presented in this section have been obtained
in the gaussian approximation.

All the physical observables that will be used in the calibration
procedure will be computed using the interaction Lagrangian density
of Ref.\,\cite{Mueller:1996pm}---properly supplemented by a mixed
isoscalar-isovector term\,\cite{Horowitz:2000xj,Todd-Rutel:2005fa}.
That is,
\begin{eqnarray}
{\mathscr L}_{\rm int} &=&
\bar\psi \left[g_{\rm s}\phi   \!-\!
         \left(g_{\rm v}V_\mu  \!+\!
    \frac{g_{\rho}}{2}{\mbox{\boldmath $\tau$}}\cdot{\bf b}_{\mu}
                               \!+\!
    \frac{e}{2}(1\!+\!\tau_{3})A_{\mu}\right)\gamma^{\mu}
         \right]\psi \nonumber \\
                   &-&
    \frac{\kappa}{3!} (g_{\rm s}\phi)^3 \!-\!
    \frac{\lambda}{4!}(g_{\rm s}\phi)^4 \!+\!
    \frac{\zeta}{4!}   g_{\rm v}^4(V_{\mu}V^\mu)^2 +
   \Lambda_{\rm v}\Big(g_{\rho}^{2}\,{\bf b}_{\mu}\cdot{\bf b}^{\mu}\Big)
                           \Big(g_{\rm v}^{2}V_{\nu}V^{\nu}\Big)\;.
 \label{LDensity}
\end{eqnarray}
The Lagrangian density includes an isodoublet nucleon field ($\psi$)
interacting via the exchange of two isoscalar mesons, a scalar
($\phi$) and a vector ($V^{\mu}$), one isovector meson (${\bf
b}^{\mu}$), and the photon
($A^{\mu}$)\,\cite{Serot:1984ey,Serot:1997xg}. The need for large
scalar ($g_{\rm s}$) and vector coupling constants ($g_{\rm v}$) is
the hallmark of the very successful relativistic mean-field
theories\,\cite{Walecka:1974qa}, that account for the saturation of
symmetric nuclear matter and the large spin-orbit splitting
displayed by finite nuclei. However, to improve the standing of the
model the Lagrangian density must be supplemented by scalar and
vector self-interactions.  In particular, cubic and quartic scalar
self-interactions (with coupling constants $\kappa$ and $\lambda$)
were first introduced by Boguta and Bodmer in
1977\,\cite{Boguta:1977xi} to soften the equation of state (EOS) of
symmetric nuclear matter around saturation density. Such a softening
is demanded by the measured distribution of isoscalar monopole
strength in medium to heavy nuclei, as these are particularly
sensitive to the incompressibility coefficient of nuclear matter.
Further, the quartic isoscalar-vector self-interaction (with
coupling constant $\zeta$) is also responsible for a softening of
the EOS of symmetric nuclear matter---but at much higher densities.
Indeed, M\"uller and Serot were able to show that by tuning $\zeta$
one can generate maximum neutron-star masses that may differ by one
solar mass, without modifying the behavior of the EOS around
saturation density\,\cite{Mueller:1996pm}. As such, $\zeta$ is
fairly insensitive to laboratory observables and must be constrained
from astrophysical observations of massive neutron
stars\,\cite{Demorest:2010bx,Antoniadis:2013pzd}. Finally, the mixed
quartic vector interaction (as described by the parameter
$\Lambda_{\rm v}$) was introduced to modify the density dependence
of symmetry energy\,\cite{Horowitz:2001ya}, which is traditionally 
stiff in relativistic models. In particular, tuning this parameter serves 
to soften the symmetry energy and has a significant impact on the
behavior of poorly constrained isovector observables, such as the
neutron-skin thickness of heavy nuclei.

The above discussion serves to indicate that the connection between
model parameters and physical observables is a complicated one.
Indeed, all bulk parameters of nuclear matter---such as the
saturation density, binding energy, incompressibility coefficient,
symmetry-energy coefficient, and slope of the symmetry energy---are
known to involve complicated combinations of the model parameters.
Moreover, as argued in the discussion following
Eq.\,(\ref{VarianceA1}), the eigenvectors of the matrix of second
derivatives also depend on a complicated linear combination of model
parameters. In particular, small eigenvalues of the matrix of second
derivatives involve linear combinations of coupling constants that
are poorly constrained by the objective function. In fact, it was
shown in Ref.\,\cite{Fattoyev:2011ns} that one (of the two) linear
combination of the isovector parameters $g_{\rho}$ and $\Lambda_{\rm
v}$ is very soft and that such linear combination is strongly
sensitive to the slope of the symmetry energy. Thus, it seems much
more natural to define the objective function $\chi^{2}({\bf p})$
directly in terms of the various bulk parameters of nuclear matter
rather than in terms of the coupling constants. This mapping has
been shown to be possible in the case of the Skyrme
interaction\,\cite{Agrawal:2005ix,Chen:2010qx,Kortelainen:2010hv}.
As we argue later in Sec.\,\ref{Results}, the same ideas are used 
here for the first time in the relativistic case. However, in order to avoid
interrupting the flow of the narrative, we only provide here a brief
summary of the central points. A detailed account of the
transformation between Lagrangian parameters and bulk parameters
will be presented in a forthcoming publication\,\cite{Chen:2014}.

In essence, of the five isoscalar parameters given in
Eq.\,(\ref{LDensity}), namely, $g_{\rm s}$, $g_{\rm v}$ $\kappa$,
$\lambda$, and $\zeta$, the first four can be uniquely determined by
specifying four bulk properties of symmetric nuclear matter: (i) the
saturation density $\rho_{{}_{\!0}}$, (ii) the binding energy per
nucleon $\varepsilon_{{}_{\!0}}$, (iii) the incompressibility
coefficient $K$, and (iv) the effective nucleon mass $M^{\ast}$ (all
at saturation density). The determination is unique because there is
a \emph{linear} transformation relating the two sets of
quantities\,\cite{Glendenning:2000}. Left undetermined in the
isoscalar sector are then $\zeta$ and the mass of the scalar meson
$m_{\rm s}$, which can not be separated from $g_{\rm s}$ in nuclear
matter. Note that the mass of the isoscalar-vector (``$\omega$'')
meson will be fixed at its experimental value of $m_{\rm
v}\!=\!782.5\,$MeV. In the case of the two isovector parameters
$g_{\rho}$ and $\Lambda_{\rm v}$, we show in Ref.\,\cite{Chen:2014}
that both of them can be determined from knowledge of the symmetry
energy $J$ and its slope $L$ at saturation density. The mass of the
isovector (``$\rho$'') meson has been fixed at its experimental
value of $m_{\rho}\!=\!763\,$MeV. In this way the objective function
$\chi^{2}({\bf p})$ depends on a vector ${\bf p}\!\equiv\!(m_{\rm
s},\rho_{{}_{\!0}},\varepsilon_{{}_{\!0}}, M^{\ast},K,\zeta,J,L)$ in
an 8-dimensional parameter space. Given a point ${\bf p}$ in such a
parameter space, the aforementioned transformation is then used to
generate the corresponding point ${\bf q}\!=\!(m_{\rm s},g_{\rm
s},g_{\rm v},g_{\rho},\kappa,\lambda, \zeta,\Lambda_{\rm v})$ in the
space of Lagrangian parameters. Once ${\bf q}$ is determined---and
thus all parameters of the Lagrangian---one can calculate all
physical observables required to evaluate the objective function.
The enormous virtue of this approach is that the objective function
is defined in terms of (mostly) physically intuitive parameters
rather than in terms of the empirical parameters of the Lagrangian
density. Finding out that a large theoretical error is associated to
one such physical parameter, say the slope of the symmetry energy
$L$, is more informative than discovering a large error in a
coupling constant, say $\Lambda_{\rm v}$. Moreover, this choice also
significantly improves the efficiency of the calibration procedure
as the range of the physical parameters is fairly well constrained.

Once the theoretical model has been defined, one must select the set
of observables that will be used in the calibration of the objective
function. To this end we have used binding energies and charge radii
of the following ten doubly-magic (or semi-magic) nuclei:
${}^{16}$O, ${}^{40}$Ca, ${}^{48}$Ca, ${}^{68}$Ni, ${}^{90}$Zr,
${}^{100}$Sn, ${}^{116}$Sn, ${}^{132}$Sn, ${}^{144}$Sm, and
${}^{208}$Pb. Experimental binding energies were obtained from the
latest 2012 atomic mass evaluation\,\cite{AME:2012} and charge radii
(except in the cases of ${}^{68}$Ni and ${}^{100}$Sn where they are
unavailable) from Ref.\,\cite{Angeli:2013}. Besides ground-state
properties, also included in the calibration are the centroid
energies of the isoscalar monopole resonance in ${}^{90}$Zr,
${}^{116}$Sn, ${}^{144}$Sm, and ${}^{208}$Pb. Experimental centroid
energies were extracted from Refs.\,\cite{Youngblood:1999,Li:2007bp,
Patel:2013uyt} which include experiments performed at both Texas
A\&M University (TAMU) and the Research Center for Nuclear Physics
(RCNP) in Osaka, Japan. We note that the centroid energy of
${}^{208}$Pb extracted from RCNP ($13.5\pm0.1$\,MeV) is considerably
lower than the one extracted from TAMU ($14.18\pm0.11$\,MeV).
Implications of this discrepancy on the ``softness of Tin" will be
addressed in a forthcoming publication. Finally, from the recent
observation of 2 solar-mass neutron
stars\,\cite{Demorest:2010bx,Antoniadis:2013pzd}, we incorporate in
the calibration of the objective function a maximum neutron star
mass of $2.1\,M_{\odot}$. The minimization of the objective function
and the extraction of the matrix of second derivatives was
implemented by employing the Levenberg-Marquardt method---that was
found to be both stable and efficient. In analogy to the successful
FSUGold parametrization\,\cite{Todd-Rutel:2005fa}, the newly
calibrated interaction has been named FSUGold\,2. A manuscript
describing in far greater detail the calibration of FSUGold\,2 is
forthcoming\,\cite{Chen:2014}.

Before displaying a few of our results, we summarize some of the
central points of the calibration of FSUGold\,2. First, we have
implemented for the first time in the relativistic approach the
transformation from empirical coupling constants to physical
parameters. That is, the objective function is defined in terms of
bulk parameters of nuclear matter rather than coupling constants.
Not only is the implementation more efficient, but the minimization
is guided by physical intuition. Second, the calibration of the
objective function relies exclusively on \emph{measurable}
quantities---that include ground-state properties of finite nuclei,
giant monopole energies, and a maximum neutron-star mass. Finally,
other than the recent work of Ref.~\cite{Erler:2012qd}, we are
unaware of any other calibration of an EDF for finite nuclei and
neutron stars.

\begin{figure}[ht]
\vspace{-0.05in}
\includegraphics[width=0.5\columnwidth,angle=0]{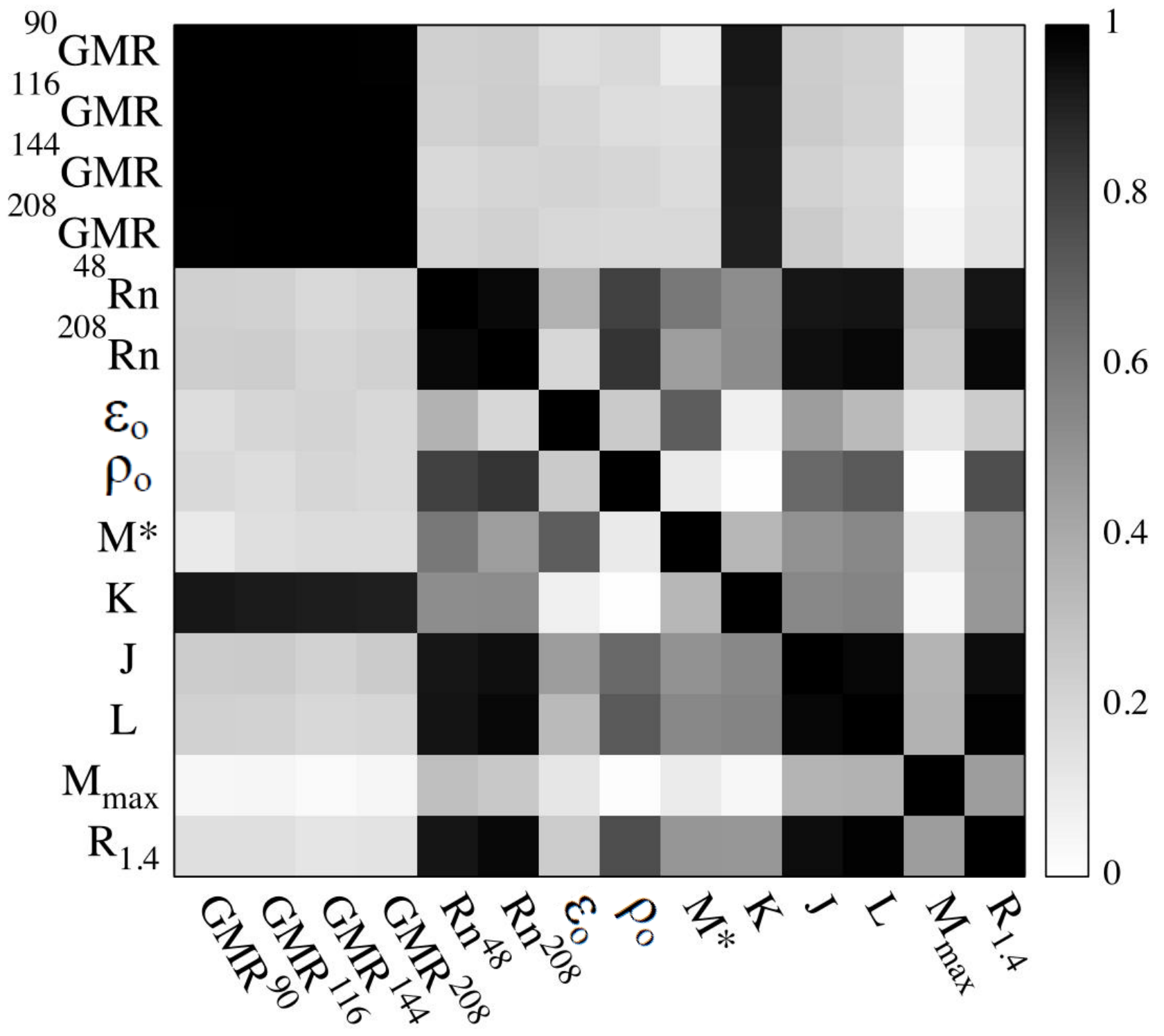}
\caption{Graphical representation of the correlation coefficients between
14 observables as predicted by the FSUGold\,2 effective interaction. The
set includes GMR centroid energies for ${}^{90}$Zr,
${}^{116}$Sn, ${}^{144}$Sm, and ${}^{208}$Pb; neutron radii for
${}^{48}$Ca and ${}^{208}$Pb; bulk properties of infinite nuclear
matter at saturation density ($\rho_{{}_{0}}$, $\varepsilon_{{}_{0}}$,
$M^{*}$, $K$, $J$, and $L$); and the maximum neutron-star mass
$M_{\rm max}$ and stellar radius $R_{1.4}$ of a 1.4
${\rm M}_{\odot}$ neutron star.}
\label{Fig3}
\end{figure}

In Fig.\,\ref{Fig3} we display in graphical form the correlation
coefficients using a set consisting of 14 observables. This set
includes GMR energies for ${}^{90}$Zr, ${}^{116}$Sn, ${}^{144}$Sm,
and ${}^{208}$Pb; neutron radii for ${}^{48}$Ca and ${}^{208}$Pb;
bulk properties of infinite nuclear matter at saturation density;
and two neutron-star properties: the maximum neutron-star mass
$M_{\rm max}$ and the radius of a 1.4 ${\rm M}_{\odot}$ neutron star
$R_{1.4}$. With the exception of the centroid energies and $M_{\rm
max}$ that were included in the calibration procedure, all the other
observables are predictions of the model. Although a thorough
analysis of these results will be presented in
Ref.\,\cite{Chen:2014}, we discuss here a few salient features.
First and as expected, all GMR energies are strongly correlated to
the incompressibility coefficient of symmetric nuclear matter $K$.
Second, we have found---as many have before us---a strong
correlation between the neutron radius of ${}^{208}$Pb and the slope
of the symmetry energy $L$. Our results suggest that such a strong
correlation persists between $R_{n}^{208}$ and $R_{1.4}$, even
though these quantities differ by 18 orders of magnitude. Finally,
the maximum neutron-star mass is weakly correlated to every
observable displayed in the figure. This suggests that, whereas
laboratory experiments can place significant constraints on
fundamental parameters of the nuclear EOS around saturation density,
the only meaningful constraint on the high-density component of the
EOS of cold neutron-rich matter must come from massive neutron
stars.

\begin{figure}[ht]
\vspace{-0.05in}
\includegraphics[width=0.45\columnwidth,angle=0]{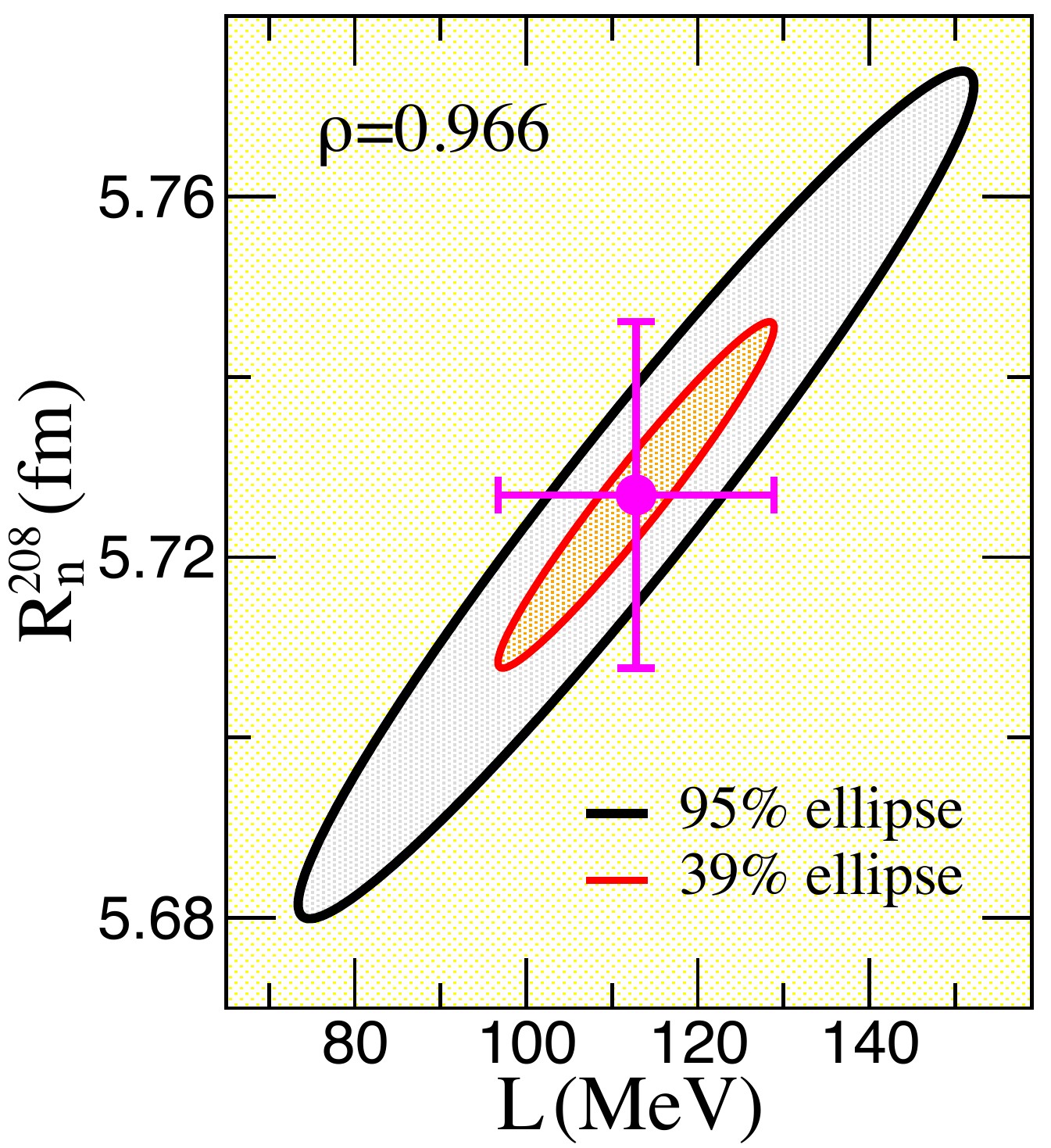}
\caption{(color online) Correlation plot between the slope of the
symmetry energy $L$ and the neutron radius of ${}^{208}$Pb
as predicted by the newly developed effective interaction
FSUGold\,2.}
\label{Fig4}
\end{figure}

We finish this section by displaying a correlation plot between the
slope of the symmetry energy and the neutron radius of $^{208}$Pb.
The newly developed relativistic EDF FSUGold\,2 predicts:
$L\!=\!(112.8\pm16.1)$\,MeV, $R_{n}^{208}\!=\!(5.727\pm0.019)$\,fm,
and a robust correlation coefficient of $\rho\!=\!0.966$. This
information has been used to produce the 39\% and 95\% confidence
ellipses displayed in Fig.\,\ref{Fig4}. In particular, our results
indicate that a 0.3\% measurement of $R_{n}^{208}$ would constrain
$L$ to about 15\%. At present, this 0.3\% requirement is beyond the
capabilities of the second phase of the Lead Radius Experiment
(PREX-II) at the Jefferson Laboratory that aims for a 1\%
measurement of $R_{n}^{208}$\,\cite{PREXII:2012}. According to our
results---by itself---a 1\% measurement of $R_{n}^{208}$ will only
be able to constrain $L$ to about 50\,MeV. However, in combinations
with other observables sensitive to the density dependence of the
symmetry energy, such as the electric dipole polarizability, one
could obtain a significantly more stringent constraint on $L$. Note
that with the above value of $R_{n}^{208}$, FSUGold\,2 predicts the
rather large neutron-skin thickness of $R_{\rm
skin}^{208}\!=\!(0.287\pm0.020)$\,fm; for comparison, FSUGold
predicts 0.207\,fm. This result indicates that as long as the
objective function does not incorporate stringent experimental
constraints on the isovector sector, relativistic models will
continue to predict large neutron
skins\,\cite{Lalazissis:1996rd,Lalazissis:1999}. Note that although
large, our result is consistent with the pioneering PREX experiment
that recently provided, albeit with large error bars, the first
model-independent evidence on the existence of a neutron-rich skin
in ${}^{208}$Pb\,\cite{Abrahamyan:2012gp,Horowitz:2012tj}:
\begin{equation}
 R_{\rm skin}^{208}\!=\!{0.33}^{+0.16}_{-0.18}\,{\rm fm}.
 \label{PREX}
\end{equation}
Moreover, our finding is also consistent with Ref.\,\cite{Fattoyev:2013yaa} 
that recently suggested that well measured physical observables as the 
ones included in this work, are unable to rule out a thick neutron skin
in ${}^{208}$Pb.

\subsection{A Word on Systematic Errors}
\label{SystErrors}

We close this section with a brief comment on the critical task of
estimating systematic errors. Although the FSUGold\,2 functional has
been accurately calibrated and statistical uncertainties were
properly computed, it is not possible to estimate the systematic
errors associated with its predictions. Given that FSUGold\,2
incorporates its own constraints, limitations, and intrinsic biases,
such systematic uncertainties can only be assessed by comparing
against the predictions of different models.

\begin{figure}[ht]
\vspace{-0.05in}
\includegraphics[width=0.45\columnwidth,angle=0]{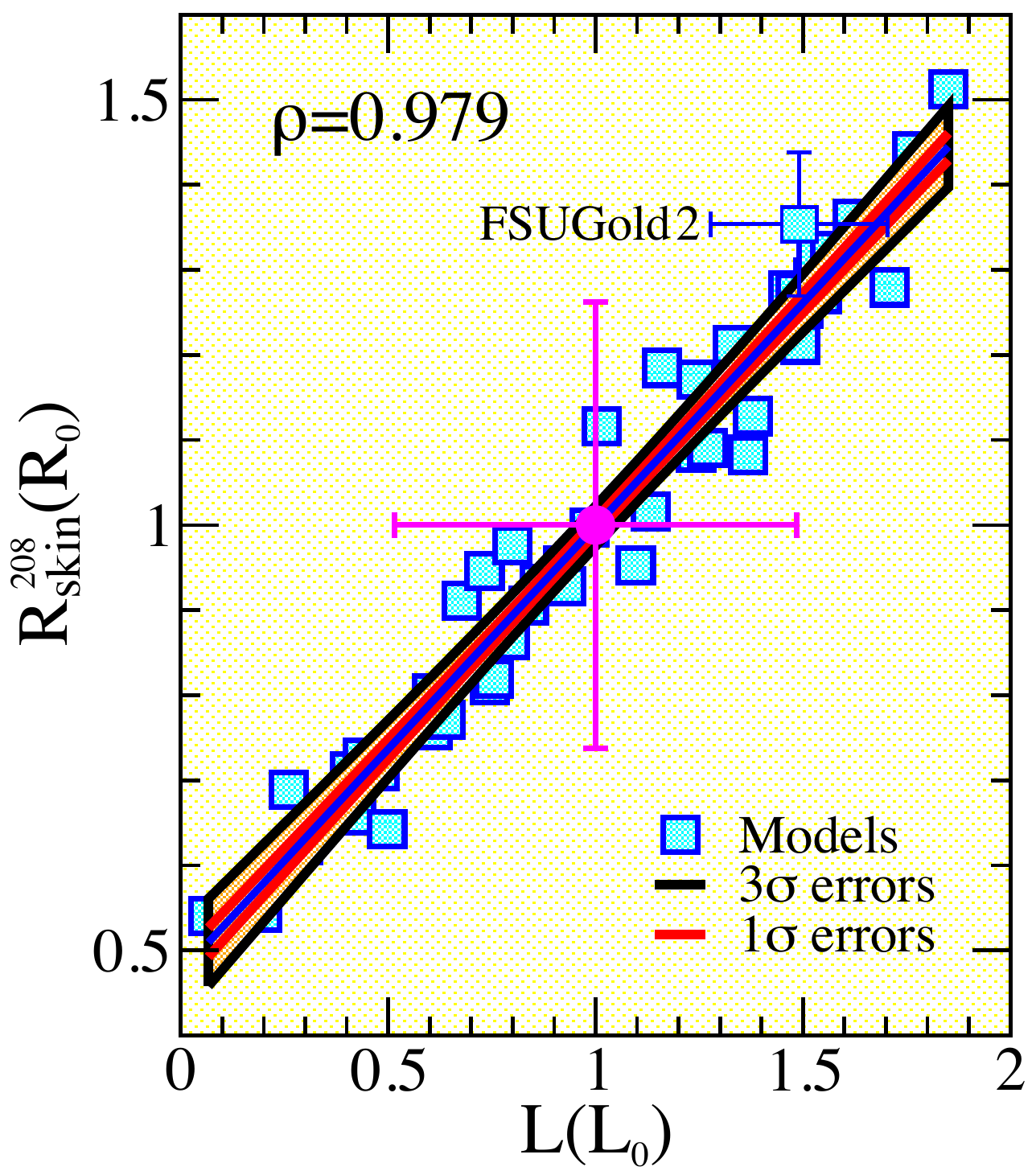}
\caption{(color online) Predictions for the slope of the symmetry
energy and the neutron-skin thickness of ${}^{208}$Pb by a large
number (47) of both relativistic and non-relativistic models in
units of the corresponding average values
$L_{0}\!\equiv\!75.676$\,MeV and $R_{0}\!\equiv\!0.212$\,fm. The
meaning of the optimal straight line and the corresponding error
bands are explained in the text. The figure was adapted from
Ref.\,\cite{RocaMaza:2011pm}.} 
\label{Fig5}
\end{figure}

To illustrate the role of systematic errors and to continue with our
current theme, we display in Fig.\,\ref{Fig5} predictions for both
the slope of the symmetry energy and the neutron-skin thickness of
${}^{208}$Pb for a large number (47) of relativistic and
non-relativistic models (this figure has been adapted from
Ref.\,\cite{RocaMaza:2011pm} with the data kindly provided by X.
Roca-Maza). The predictions from FSUGold\,2 with its associated
theoretical errors are also displayed in the figure. Here
$L_{0}\!\equiv\!75.676$\,MeV and $R_{0}\!\equiv\!0.212$\,fm
represent the average values of all 47 predictions. Note that the
dispersion in the corresponding average values are $\Delta
L\!=\!36.730$\,MeV and $\Delta R\!=\!0.055$\,fm, respectively. These
values are depicted by the (magenta) cross in the middle of the
figure. By performing a standard least-squares fit to the
predictions of the 47 models, we obtain the following optimal
straight line:
\begin{equation}
 \left(\frac{R_{\rm skin}^{208}}{R_{0}}\right) =
 b + m\left(\frac{L}{L_{0}}\right)\;,
 \label{LSFit}
\end{equation}
with a large correlation coefficient of $\rho\!=\!0.979$, and
(dimensionless) slope and intercept given by $m\!=\!(0.524\pm0.016)$
and $b\!=\!(0.475\pm0.018)$, respectively. This optimal straight
line is displayed by the blue solid line in Fig.\,\ref{Fig5}. Note
that the probability distribution of ``straight lines" $P(m,b)$ was
obtained using maximum-likelihood estimates exactly as before.
Indeed, $P(m,b)$ satisfies the same gaussian expression given in
Eqs.\,(\ref{PofAB2}) and\,(\ref{CovMtrx}) for the case of any two
generic variables\,\cite{Bevington2003}. In the present case,
however, the gaussian approximation is exact. Knowledge of the
probability distribution $P(b,m)$ not only provides an estimate of
the systematics uncertainties in the slope and intercept of the
optimal straight line but, in addition, permits to assess the
systematic errors in $R_{\rm skin}^{208}$ for any given value of
$L$. Doing so results in the $1\sigma$ (in red) and $3\sigma$ (in
black) error bands displayed in Fig.\,\ref{Fig5}. That is, the 47
models given in the figure suggest that a ``new" model with a value
of, for example, $L\!=\!1.5L_{0}$ (as FSUGold\,2) will predict a
neutron-skin thickness for ${}^{208}$Pb that will have a 68.3\%
probability of falling within the interval $R_{\rm
skin}^{208}\!=\!(1.26\pm0.01)R_{0}$. Besides being useful in
providing an estimate of the systematic errors, plots as the one
given in Fig.\,\ref{Fig5} are useful in quantifying how accurately a
measurement (of for example $R_{\rm skin}^{208}$) should be done in
order to rule out some theoretical models. Of course, deciding
whether a model can be ruled out or ruled in depends critically on
the \emph{statistical} uncertainties predicted by such model, which
emphasizes again the importance of providing theoretical
uncertainties.

\section{Conclusions and Outlook}
\label{Conclusions} As aptly captured by the Editors of the Physical
Review A\,\cite{PhysRevA.83.040001}, the central goal of this
contribution to the focus issue on information and statistics is
\emph{`to discuss the importance of including uncertainty estimates
in papers involving theoretical calculations of physical
quantities'}. Given the impossibility of first-principle
calculations of nuclear phenomena using QCD, theoretical nuclear
physics must rely on the modeling of such phenomena. In this regard,
the most sophisticated theoretical framework is density functional
theory. Whereas in principle the empirical constants defining the
EDF should be computed from QCD, in practice they are fitted
directly to many-body observables. As such, the calibration of the
EDF relies on the minimization (or calibration) of a properly
defined objective function. Until recently, once the minimum was
found the model was validated against observables not included in
the fit. More recently, however, the standards have been raised
considerably. As properly articulated in
Ref.\,\cite{PhysRevA.83.040001}, one now demands predictions
to be accompanied by meaningful and reliable theoretical errors.

In this contribution we have framed the formalism in terms of
maximum-likelihood estimation. Such an approach is intuitive,
transparent, and---at least in principle---straightforward to
implement. Once a model has been selected and a set of
accurately-measured observables identified, the objective function
(e.g., $\chi^{2}$) is defined in terms of the square-differences
between these observables and the predictions of the model. The
likelihood function---simply obtained from ``exponentiation" of the
objective function---now acts as an un-normalized probability
distribution. In this manner, one could generate, for example via
Monte-Carlo methods, the distribution of models in the vast space of
parameters. The maximum-likelihood estimate provides the optimal set
of parameters, namely, the minimum of the objective function.
However, maximum-likelihood estimation goes well beyond providing
the optimal model. Indeed, by generating model parameters according
to the likelihood function, one can now estimate theoretical errors
and correlations among any pair of observables using conventional
statistical averaging.

Given its very modest computational demands, we have used the
revered liquid-drop model to illustrate the power and elegance of
maximum-likelihood estimation (see also
Ref.\,\cite{Dobaczewski:2014jga}). By doing so---not only can we
obtain theoretical averages and errors, but in addition---we were
able to generate the full probability distribution of the
liquid-drop parameters. Moreover, we could assess the degree of
correlation among the parameters through highly insightful
confidence ellipses. Finally, we relied on this pedagogical model to
test the robustness of the \emph{gaussian approximation}, namely,
the approximation in which the deviations from the $\chi^{2}$
minimum are assumed to be quadratic.

For the more realistic case of the optimization of a relativistic
EDF, we were limited---because of severe computational demands---to
the gaussian approximation. Once the relativistic Lagrangian was
defined and physical observables selected (which involve
ground-state properties of finite nuclei, giant-monopole energies,
and the limiting mass of a neutron star) the minimum of the
objective function and its associated matrix of second derivatives
were obtained using the Levenberg-Marquardt method. With such
information at hand, we were able to compute averages, theoretical
errors, and correlation coefficients for various observables. In
particular, we concluded that the maximum mass of a neutron star is
poorly correlated to a host of laboratory observables---suggesting
that the search for massive neutron stars may provide the only
meaningful constraint on the EOS of neutron-star matter. A more
detailed discussion of these results will be presented in a
forthcoming publication\,\cite{Chen:2014}. We note that the
minimization procedure was implemented by using, for the first time
in the relativistic approach, a parameter set consisting of physical
quantities rather than empirical parameters of the Lagrangian. The
transformation from empirical constants to physical quantities
provides several important advantages\,\cite{Chen:2014}.

In summary, we have used central idea from information and
statistics---particularly maximum-likelihood estimation---to
demonstrate the power of the approach and its critical role in the
development of a new paradigm in nuclear theory. The need to
quantify theoretical uncertainties is particularly urgent as models
that are fitted to experimental data are then used to extrapolate to
the extremes of density and isospin asymmetry, such as those
encountered in neutron stars. The present focus issue in
general---and this contribution in particular---provide detailed
derivations and simple examples on how to apply these ideas to the
quantification of uncertainties and the analysis of correlations.
Besides the obvious benefits, a proper statistical analysis may also
improve the control of the systematic errors. Although there are
many sources of systematic errors, one that may be readily addressed
involves comparison among various accurately-calibrated models.
Given that models include intrinsic constraints and limitations,
their prediction may vary significantly---even when calibrated to
the same set of experimental data. Such systematic uncertainties can
only be assessed by comparing different models. However, the
predictions from the various models are rarely (if ever) accompanied
by their corresponding statistical uncertainties. Hence, it is
common practice to simply ``throw" all predictions into a plot and
then attempt to infer some degree of correlation among the
observables of interest. This situation is unsuitable and one should
advocate for higher quality control involving, at the very least,
theoretical predictions with properly quantified uncertainties. It
appears that the contributors to the Physical Review A have adapted well
to such higher demands. It is high time for the nuclear-theory
community to follow suit.

\begin{acknowledgments}
\vspace{-0.4cm} We thank the editors for inviting us to contribute
to this focus issue. This work was supported in part by grant
DE-FD05-92ER40750 from the United States Department of Energy, by
the National Aeronautics and Space Administration under grant
NNX11AC41G issued through the Science Mission Directorate, and the
National Science Foundation under grant PHY-1068022.
\end{acknowledgments}
\vfill
\bibliography{JPhysG_Final.bbl}
\vfill\eject
\end{document}